# Forecasting Global Network Traffic Trends: The Role of Virtual Reality

First . Raghad H.AlShekh, Second Qutaiba Ali *Member, IEEE*

*Abstract*—Virtual Reality (VR) technology demands real-time data transmission to deliver an immersive and interactive user experience. This study investigates the implementation of UDP Ethernet communication in VR systems, focusing on its impact on network performance. Experiments were conducted to analyze how factors such as cable length, data rate, and packet processing rate (PPR) influence system performance. A series of tests were performed, and the results were visualized through detailed graphs. The findings reveal how variations in these parameters affect communication speed and stability, providing insights for optimizing VR system design. By leveraging the high-speed, low-overhead advantages of UDP Ethernet, this study may contribute understanding of network performance in VR applications, offering practical guidance for developers and engineers in creating responsive and efficient VR environments.

*Index Terms*— Emerging technologies Virtual Reality, Modelling, Performance.

## I. INTRODUCTION

The rapid evolution of emerging technologies is transforming the way we interact with digital content, leading to an unprecedented rise in network traffic demands. These innovations are introducing new complexities to network infrastructures, driven by high data rates, low-latency requirements, and the integration of advanced applications [1]. To ensure that networks can meet these future demands, it is essential to develop frameworks that estimate traffic requirements [2].

This research focuses on anticipating the impact of new technologies on network performance and infrastructure planning. Through the analysis of technology trend data, traffic modeling, and expert insights, this work provides valuable tools for adapting to the evolving landscape of network requirements while supporting the seamless deployment of technologies

The concept of **virtual reality (VR)** emerged in the late '80s and '90s, but it had its origins in the 1950s. Specifically, it is a three-dimensional simulation of an artificial environment that has the potential to provide users with experiences that are either similar to the actual world or completely distinct from it [3]

Tom Caudell developed augmented reality (**AR**) in 1990. It enhances the real-time environment or physical world by superimposing interactive virtual 3D features that resemble real-life entities [4].

Mixed Reality (MR) refers to the merging of VR and AR technologies to create a hybrid environment and combine virtual and actual surroundings. It creates an environment with interactive digital objects where the line between virtual and physical reality is blurred [5].

The **Metaverse** known as a large-scale, persistent, and immersive network of 3D virtual environments, seamlessly generates digital replicas of real spaces called digital twins, which can be used to simulate scenarios and optimizations before implementing them in the real world [6]. It is facilitated by a convergence of various technologies, such as AR, MR, and VR [7] As shown in figure (1)

Those technologies have found extensive use in several domains medical, cultural, product development, and online shopping,etc. [8].

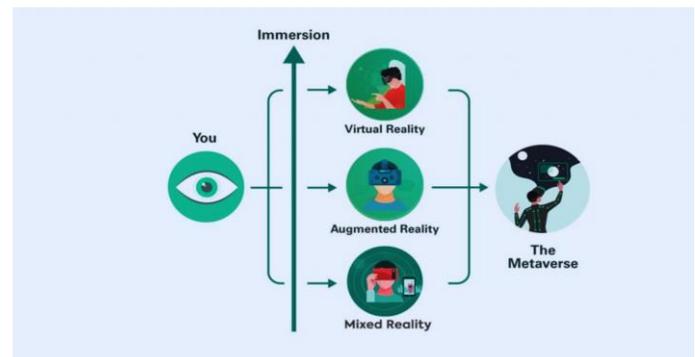

Fig .1. Trended technologies

The paper is organizing as follows: Section II shows the literature review. Section III presents a case study taken on VR technology performance. The results and discussion is described in Section IV. Finally, section V. summarizes the paper and suggests future works.

## II. LITERATURE REVIEW

Table (I) at of this research summarizes thirteen research papers on different immersive technologies, with each paper representing a specific problem description, contribution, and outcomes.

> REPLACE THIS LINE WITH YOUR PAPER IDENTIFICATION NUMBER (DOUBLE-CLICK HERE TO EDIT) <       2Table (I) Literature Review Summary

| Pager | Problem statement | Objective | Tools | Contribution | Results |
|---|---|---|---|---|---|
| Himanshu et al. 2020 [9] | VR enhance training in health care and identify immersion breakpoints. | Explain VR in healthcare, AI integration, identify Immersion Breakpoints | VRTK, blender, Maya. | Remote training and therapy options. | Realistic simulations for surgical practice. Effective treatment for psychological conditions. |
| Fengxian Guo et al. 2020 [10] | Ignoring real-time rendering and data correlation | catch MEC capability, performance improvement, Maximize QOE | Deep reinforcement learning DRL, game theory | Enhance QOE, introduce distributed learning algorithm | outperforms the baseline algorithms in terms of QoE, latency, and convergence time |
| LIDIA M. ORTEGA et al 2020 [11] | Underground and indoor facilities lack real-time data and rely on outdated plans | manage 3D underground infrastructure networks | 3D GIS and BIM, CityGML | topological data model | Successful execution of CRUD operations |
| Venkatakrishnan et al 2020 [12] | Multi-user VR needs high network performance, QoS, privacy, and high image quality. | Evaluate multi-user VR application network | Oculus Rift, headset, Python | analyzes network conditions and user counts on VR performance. | More than 10 users degrade performance, improve smoothness |
| A. Hayes, et al 2021 [3] | limitations to implementing VR in education. | enhance 3D VR Learning Environments (VRLE). | ----- | introduces TPACK framework | supporting the technological competency of K16 teachers, integrate VR content into classrooms effectively |
| Chufeng Huang et al 2022 [13] | Insufficient data sharing and openness | propose a virtual reality scene modeling method that integrates IoT technology. | edge computing, database dynamic loading management | large-scale data acquisition and modeling | Improved efficiency, Enhanced capabilities |
| Dimitrios Ververidis et al 2022 [14] | AEC industry needs better VR and BIM integration | state-of-the-art collaborative VR systems | Building Information Modeling, Computer-Aided Design | proposed guidelines for collaborative VR system. | proposed a blueprint for an ideal system. |
| Saeed Safikhani et al., [15] | AEC industry struggles to use BIM fully due to collaboration challenges | Identify VR-BIM use cases | PRISMA was conducted VR and BIM terms. | Design and project management, and provides insights into VR-BIM implementation. | improves understanding, decision-making, and collaboration. |
| Kai Zhang et al., 2023 [16] | accurate and real-time network traffic prediction, | Spatial-Temporal Graph Convolution Gated Recurrent Unit (GC-GRU) | GC-GRU Model | model (GC-GRU) | handling changing data and working well in various situations. |
| Francesca Bruni et al., 2024 [17] | create program using semi-immersive VR | Integrate VR with ML to help Parkinson's sick people | Cave Automatic Virtual Environment, (Full HD 3D UXGA DLP, dual-Task Exercises | establish a novel approach to dual-task rehabilitation that combines cognitive and motor training in realistic manner. | increase treatment's and individualized training. |
| Adedotun Adetunla 2024 [18] | Challenges of VR in education | integration of VR into educational settings, enhance teaching | headsets | Highlighting diverse applications of VR in education | VR enhances student motivation |

## III. CASE STUDY: PREDICTING VR NETWORK DEMANDS

Virtual Reality (VR) networking plays a strategic role in guaranteeing continuous and engaging user interfaces. Bear in mind that in order to get the best out of VR, a number of network parameters, latencies as well as bandwidth are critical in the face of high data rates and low latency requirements [19].

### A. Case description

The main objective of this case study is to develop a framework to estimate the traffic demands of VR, by considering factors like data rate, Packet processing rate, and cable length to avoid potential bottlenecks. The network consists of VR client and server connected via Ethernet cable. As shown in figure (2).

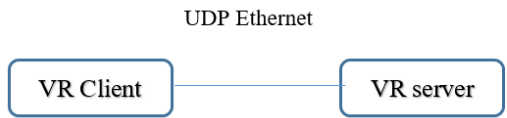

Fig .2. system block diagram

### B. Research method

Develop a virtual reality model to test network behavior under a variety of application and network constraints. Analyze variables such as cable length, data rates, and packet processing rate (PPR) to determine how they affect network performance.

### C. Modelling implementation

Suggest a VR system that uses UDP Ethernet network with specifications shown on table (II), notice that all sizes are in bytes.

Table (II) Suggested VR network specifications [18], [20]–[22]

| Parameter | Value | Parameter | value |
|---|---|---|---|
| Ethernet frame header | 14 | MTU | 1500 |
| IP header | 20 | VR fps | 60 |
| UDP header | 8 | Max payload size | 1432 |
| VR Protocol base header | 26 | VR frame size | 1 MB |

The system is described as a comprehensive sequence diagram that illustrates the entire communication flow for VR applications using UDP Ethernet networks between a VR Client and a VR Server during a virtual reality session. The sequence is divided into three major phases: Session Establishment phase, Active Video Streaming phase in which the actual video streaming and interaction between the client and server take place, and Session Termination phase. Each phase contains several necessary signals to finish the session. In addition to these three basic phases, there are more phases. Shown in figure (3) To analyze the impact of each parameter on performance, we applied the system equations as shown below. Then, set two parameters and change the third to obtain graphs illustrating their effects on latency and throughput.

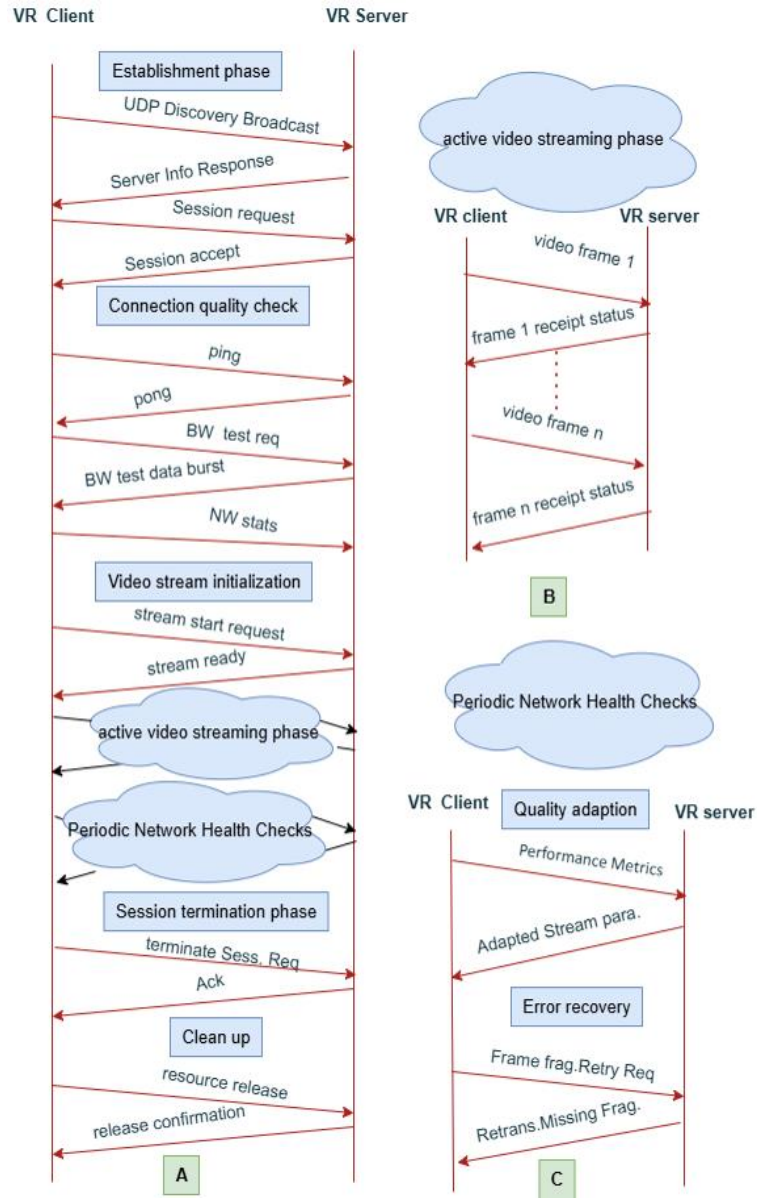

Fig .3. VR Sequence Diagram (a,b,c)

Assume periodic network health check is done every 50 ms, cable length is 10 m, data rate is 1GB and ppr is equal to 205k, and by combining stream signals and control signals in the sequence diagram in figure(3) we found that the mentioned signals impact on network latency does not exceed 0.19%, therefore we will neglect them when calculating whole system performance.

$$Maximum\ payload\ size = Ethernet\ maximum\ transfer\ unit - headers \quad ….equ.(1)$$

$$Max.payload\ size = MTU - IP - UDP - VR\ app. = 1500 - 14 - 20 - 8 - 26 = 1432 bytes$$

$$Total\ latency = Ethernet\ data\ latency + Ack.latency \quad \ldots equ.(2)$$

$$Throughput = \frac{ethernet\ data\ frame * 8}{total\ latency} \quad \ldots equ.(3)$$

$$Ethernet\ data\ latency = Ethernet\ data\ network\ delay + switch\ delay + \frac{1}{packet\ processing\ rate} \quad \ldots equ.(4)$$

$$Ack.latency = network\ delay\ Ack. + swich\ delay + \frac{1}{packet\ processing\ rate} \quad \ldots equ.(5)$$

$$Ethernet\ data\ network\ delay = \frac{data\ packet\ length}{data\ rate} + \frac{cable\ length}{propagation\ speed} \quad \ldots equ.(6)$$

$$no.of\ ethernet\ frame\ for\ each\ VR\ frame = \frac{VR\ frame\ size}{Max\ payload\ size} = \frac{IMB}{1432} = 733 \quad \ldots equ.(7)$$

$$Network\ Utilization = \frac{(data\ packet\ length + ack\ packet\ length) * (\frac{VR\ frame\ size}{Ethernet\ frame\ size}) * (VR\ fps * 8)}{data\ rate} \quad \ldots equ.(8)$$

By applying these equations to the network specifications. With a switch delay of 1 microsecond, a propagation delay of $2*10^8$ m/sec, a data packet length of 1500 bytes, and adjusting the ppr, cable length, and data rate, we obtained the results shown below.
Results and Discussion
Increasing the number of packets processing rate enhances the system's throughput. This is likely because processing larger batches of data reduces overhead per packet.
Sending more packets in each round likely reduces the frequency of handshakes or re initializations, minimizing delays and optimizing transmission.as shown in figure (4)

Latency increases linearly with cable length, starting from ~20 microseconds for short cables (0-200 meters) and rising to ~35 microseconds at 1000 meters. Throughput decreases as the cable length increases, starting from ~0.60 Gbps at short cable lengths (0-200 meters) and dropping to ~0.40 Gbps at 1000 meters. The reduction in throughput is likely due to signal degradation and increased error rates over longer cable lengths, which necessitate retransmissions and error correction mechanisms , as shown in figure (5)

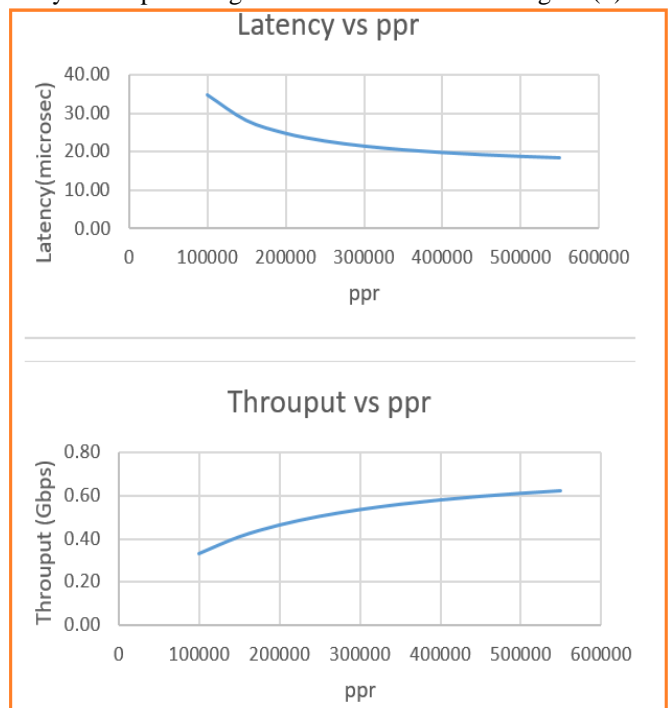

Fig .4. Latency and Throughput and  vs ppr

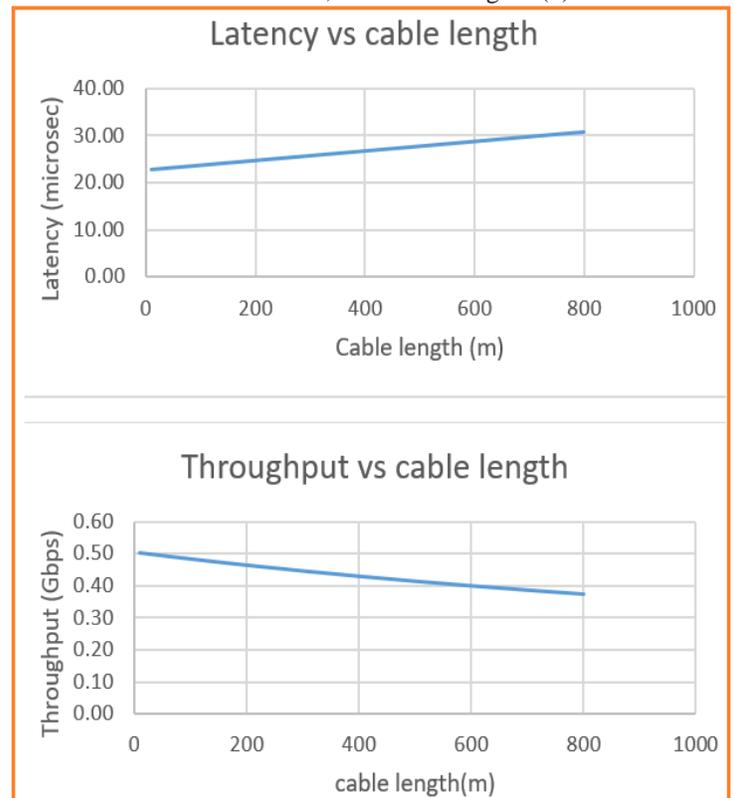

Fig .5. Latency and Throughput vs cable length

In figure (6) The throughput increases steadily as the data rate rises, showing a nearly linear trend up to the highest data rate tested (~2.5 Gbps). At higher data rates, the curve may show signs of saturation (flattening).The increase in throughput is expected since higher data rates allow for more bits to be transferred per second. The latency decreases sharply as the data rate increases, stabilizing at lower latency values (~20 microseconds) as the data rate ( ~2.5 Gbps).

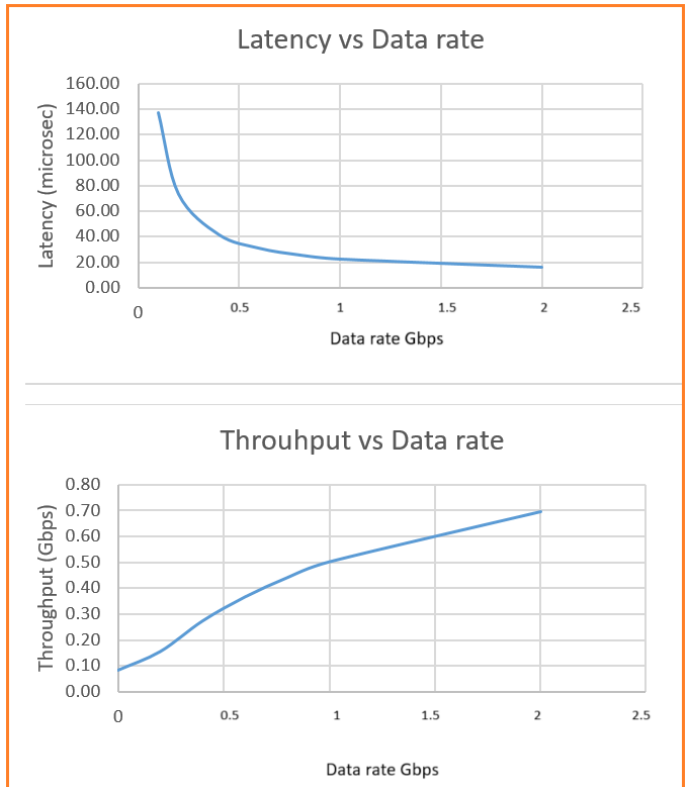

Fig .6.latency and Throughput vs data rate

### IV. TRENDED TECHNOLOGIES CHALLENGES

The advancement of technologies encounters several challenges [23], [24]. Figure 7 outlines the key obstacles faced by these technologies [5].

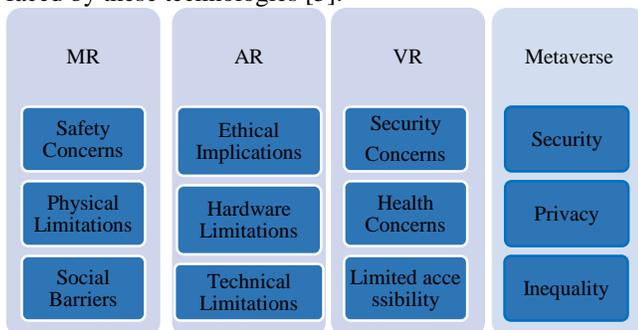

FIG .7. TRENDED TECHNOLOGIES CHALLENGES

### V. CONCLUSION

Integrating virtual reality (VR) into network design marks a significant shift in information dissemination and acquisition, as well as keeping up with the rapid and vast advancements of modern technology.

This paper investigates the impact of emerging technologies on network performance, focusing on Virtual Reality (VR) as a case study. This study highlights the critical role of UDP Ethernet communication in enhancing the performance of Virtual Reality (VR) systems. The experiments conducted demonstrate that key network parameters—such as cable length, data rate, and packets per refresh—significantly influence both latency and throughput. The findings indicate that while increasing the number of packets can improve throughput, it also introduces potential bottlenecks due to system limitations. The results underscore the importance of optimizing network configurations to meet the high demands of immersive technologies. New technologies, such as VR, hold promise and can benefit a variety of industries, including gaming, commerce, and entertainment. However, due to the huge volume of data it creates, networks must be developed, and hardware must satisfy the new criteria, such as high ppr, high processing speed, and privacy concerns.

In future Investigate the integration of other communication protocols alongside UDP to assess their impact on VR performance, particularly under varying network conditions. Exploration of the interaction between VR and other emerging technologies such as Augmented Reality (AR) and Mixed Reality (MR) to develop comprehensive solutions for multi-faceted immersive environments.

**Raghad H. AlShekh** received the B.Sc. degree and M.S. degrees from the Computer Engineering Departement from University of Mosul, Iraq. in 2004,2021 respectively, in  Currently, she is working for PhD Degree in Computer Engineering at Mosul University's College of Engineering. She has been employed at the Ministry of Ministry of Higher Education and Scientific Research in Iraq since 2008 until now; she is interested in doing research in Internet of Things, Network, Cloud Computing .

**Qutaiba I. Ali:** received the B.S. and M.S. degrees from the Department of Electrical Engineering, University of Mosul, Iraq, in 1996 and 1999, respectively. He received his Ph.D. degree (with honour) from the Computer Engineering Department, University of Mosul, Iraq, in 2006. Since 2000, he has been with the Department of Computer Engineering, Mosul University, Mosul, Iraq, where he is currently a lecturer. His research interests include computer networks analysis and design, real time networks and systems, embedded network devices and network security and managements. Dr. Ali has attended and participates in many scientific activities and gets awards for his active contribution